\font\tenfrakturb=eufb10
\font\tenfraktur=eufm10
\font\tenmsbm=msbm10
\font\sevenfrakturb=eufb7
\font\sevenfraktur=eufm7
\font\sevenmsbm=msbm7
\font\fivefrakturb=eufb5
\font\fivefraktur=eufm5
\font\fivemsbm=msbm5
\newfam\bgothicfam
\newfam\gothicfam
\newfam\msbmfam
\textfont\bgothicfam = \tenfrakturb \scriptfont\bgothicfam=\sevenfrakturb
\scriptscriptfont\bgothicfam=\fivefrakturb

\textfont\gothicfam = \tenfraktur \scriptfont\gothicfam=\sevenfraktur
\scriptscriptfont\gothicfam=\fivefraktur

\textfont\msbmfam = \tenmsbm \scriptfont\msbmfam=\sevenmsbm
\scriptscriptfont\msbmfam=\fivemsbm

\def\frak{\tenfraktur\fam\gothicfam}
\def\Bbb{\tenmsbm\fam\msbmfam}

\catcode`@=11
\def\renewcounter#1{\@definecounter{#1}\@ifnextchar[{\@newctr{#1}}{}}

\documentstyle[twoside]{article}
\catcode`\@=11
\long\def\@makefntext#1{
\protect\noindent \hbox to 3.2pt {\hskip-.9pt
$^{{\eightrm\@thefnmark}}$\hfil}#1\hfill} 

\def\@makefnmark{\hbox to 0pt{$^{\@thefnmark}$\hss}} 

\def\ps@myheadings{\let\@mkboth\@gobbletwo
\def\@oddhead{\hbox{}
\rightmark\hfil\eightrm\thepage}
\def\@oddfoot{}\def\@evenhead{\eightrm\thepage\hfil
\leftmark\hbox{}}\def\@evenfoot{}
\def\sectionmark##1{}\def\subsectionmark##1{}}
\oddsidemargin=\evensidemargin
\newcounter{sectionc}\newcounter{subsectionc}\newcounter{subsubsectionc}
\renewcommand{\section}[1] {\vspace{12pt}\addtocounter{sectionc}{1}
\setcounter{subsectionc}{0}\setcounter{subsubsectionc}{0}\noindent
	{\tenbf\thesectionc. #1}\par\vspace{5pt}}
\renewcommand{\subsection}[1] {\vspace{12pt}\addtocounter{subsectionc}{1}
	\setcounter{subsubsectionc}{0}\noindent
	{\bf\thesectionc.\thesubsectionc. {\kern1pt \bfit #1}}\par\vspace{5pt}}
\renewcommand{\subsubsection}[1] {\vspace{12pt}\addtocounter{subsubsectionc}{1}
	\noindent{\tenrm\thesectionc.\thesubsectionc.\thesubsubsectionc.
	{\kern1pt \tenit #1}}\par\vspace{5pt}}
\newcommand{\nonumsection}[1] {\vspace{12pt}\noindent{\tenbf #1}
	\par\vspace{5pt}}
\newcounter{appendixc}
\newcounter{subappendixc}[appendixc]
\newcounter{subsubappendixc}[subappendixc]
\renewcommand{\thesubappendixc}{\Alph{appendixc}.\arabic{subappendixc}}
\renewcommand{\thesubsubappendixc}
	{\Alph{appendixc}.\arabic{subappendixc}.\arabic{subsubappendixc}}
\renewcommand{\appendix}[1] {\vspace{12pt}
        \refstepcounter{appendixc}
        \setcounter{figure}{0}
        \setcounter{table}{0}
        \setcounter{lemma}{0}
        \setcounter{theorem}{0}
        \setcounter{corollary}{0}
        \setcounter{definition}{0}
        \setcounter{equation}{0}
        \renewcommand{\thefigure}{\Alph{appendixc}.\arabic{figure}}
        \renewcommand{\thetable}{\Alph{appendixc}.\arabic{table}}
        \renewcommand{\theappendixc}{\Alph{appendixc}}
        \renewcommand{\thelemma}{\Alph{appendixc}.\arabic{lemma}}
        \renewcommand{\thetheorem}{\Alph{appendixc}.\arabic{theorem}}
        \renewcommand{\thedefinition}{\Alph{appendixc}.\arabic{definition}}
        \renewcommand{\thecorollary}{\Alph{appendixc}.\arabic{corollary}}
        \renewcommand{\theequation}{\Alph{appendixc}.\arabic{equation}}
        \noindent{\tenbf Appendix \theappendixc #1}\par\vspace{5pt}}
\newcommand{\subappendix}[1] {\vspace{12pt}
        \refstepcounter{subappendixc}
        \noindent{\bf Appendix \thesubappendixc. {\kern1pt \bfit #1}}
	\par\vspace{5pt}}
\newcommand{\subsubappendix}[1] {\vspace{12pt}
        \refstepcounter{subsubappendixc}
        \noindent{\rm Appendix \thesubsubappendixc. {\kern1pt \tenit #1}}
	\par\vspace{5pt}}
\newcommand{\textlineskip}{\baselineskip=13pt}
\newcommand{\smalllineskip}{\baselineskip=10pt}
\def\eightcirc{
\begin{picture}(0,0)
\put(4.4,1.8){\circle{6.5}}
\end{picture}}
\def\eightcopyright{\eightcirc\kern2.7pt\hbox{\eightrm c}}
\newcommand{\copyrightheading}[1]
	{\vspace*{-2.5cm}\smalllineskip{\flushleft
	{\footnotesize Modern Physics Letters A #1}\\
	{\footnotesize $\eightcopyright$\, World Scientific Publishing
	 Company}\\
         }}
\newcommand{\pub}[1]{{\begin{center}\footnotesize\smalllineskip
	Received #1\\
	\end{center}
        }}

\def\abstracts#1#2#3{{
        \centering{\begin{minipage}{4.5in}\baselineskip=10pt\footnotesize
        \parindent=0pt #1\par
        \parindent=15pt #2\par
        \parindent=15pt #3
        \end{minipage}}\par}}

\newcommand{\bibit}{\nineit}
\newcommand{\bibbf}{\ninebf}
\renewenvironment{thebibliography}[1]
         {\frenchspacing
         \ninerm\baselineskip=11pt
         \begin{list}{\arabic{enumi}.}
         {\usecounter{enumi}\setlength{\parsep}{0pt}
         \setlength{\leftmargin 12.7pt}{\rightmargin 0pt} 
         \setlength{\itemsep}{0pt} \settowidth
         {\labelwidth}{#1.}\sloppy}}{\end{list}}
\newcounter{itemlistc}
\newcounter{romanlistc}
\newcounter{alphlistc}
\newcounter{arabiclistc}

\newcommand{\fcaption}[1]{
         \refstepcounter{figure}
         \setbox\@tempboxa = \hbox{\footnotesize Fig.~\thefigure. #1}
         \ifdim \wd\@tempboxa > 5in
           {\begin{center}
         \parbox{5in}{\footnotesize\smalllineskip Fig.~\thefigure. #1}
            \end{center}}
        \else
             {\begin{center}
             {\footnotesize Fig.~\thefigure. #1}
              \end{center}}
        \fi}
\newcommand{\tcaption}[1]{
        \refstepcounter{table}
        \setbox\@tempboxa = \hbox{\footnotesize Table~\thetable. #1}
        \ifdim \wd\@tempboxa > 5in
           {\begin{center}
        \parbox{5in}{\footnotesize\smalllineskip Table~\thetable. #1}
            \end{center}}
        \else
             {\begin{center}
             {\footnotesize Table~\thetable. #1}
              \end{center}}
        \fi}
\def\@citex[#1]#2{\if@filesw\immediate\write\@auxout
        {\string\citation{#2}}\fi
\def\@citea{}\@cite{\@for\@citeb:=#2\do
        {\@citea\def\@citea{,}\@ifundefined
        {b@\@citeb}{{\bf ?}\@warning
        {Citation `\@citeb' on page \thepage \space undefined}}
        {\csname b@\@citeb\endcsname}}}{#1}}
\newif\if@cghi
\def\cite{\@cghitrue\@ifnextchar [{\@tempswatrue
        \@citex}{\@tempswafalse\@citex[]}}
\def\citelow{\@cghifalse\@ifnextchar [{\@tempswatrue
        \@citex}{\@tempswafalse\@citex[]}}
\def\@cite#1#2{{$\null^{#1}$\if@tempswa\typeout
        {IJCGA warning: optional citation argument
        ignored: `#2'} \fi}}

\def\pmb#1{\setbox0=\hbox{#1}
        \kern-.025em\copy0\kern-\wd0
        \kern.05em\copy0\kern-\wd0
        \kern-.025em\raise.0433em\box0}

\def\fnt#1#2{\footnotetext{\kern-.3em
        {$^{\mbox{\scriptsize #1}}$}{#2}}}
\def\fpage#1{\begingroup
\voffset=.3in
\thispagestyle{empty}\begin{table}[b]\centerline{\footnotesize #1}
       \end{table}\endgroup}
\def\runninghead#1#2{\pagestyle{myheadings}
\markboth{{\protect\footnotesize\it{\quad #1}}\hfill}
{\hfill{\protect\footnotesize\it{#2\quad}}}}
\font\tenrm=cmr10
\font\tenit=cmti10
\font\tenbf=cmbx10
\font\bfit=cmbxti10 at 10pt
\font\ninerm=cmr9
\font\nineit=cmti9
\font\ninebf=cmbx9
\font\eightrm=cmr8

\def\bh{${\Bbb R}^2\times {\Bbb S}^2\>$}
\def\qed{\hbox{${\vcenter{\vbox{  
   \hrule height 0.4pt\hbox{\vrule width 0.4pt height 6pt
   \kern5pt\vrule width 0.4pt}\hrule height 0.4pt}}}$}}

\begin{document}
\runninghead{Yu. P. Goncharov}
{U(N)-Monopoles on Kerr Black Hole and Its Entropy}

\normalsize\textlineskip
\thispagestyle{empty}
\setcounter{page}{1}
\copyrightheading{}
\vspace*{0.88truein}
\fpage{1}
\centerline{\bf U(N)-MONOPOLES ON KERR BLACK HOLE}
\vspace*{0.035truein}
\centerline{\bf AND ITS ENTROPY
\vspace*{0.035truein}
\footnote{PACS Nos.: 04.20.Jb, 04.70.Dy, 14.80.Hv}}
\vspace*{0.37truein}
\centerline{\footnotesize YU. P. GONCHAROV}
\vspace*{0.015truein}
\centerline{\footnotesize\it Theoretical Group,
Experimental Physics Department, State Technical University}
\baselineskip=10pt
\centerline{\footnotesize\it Sankt-Petersburg 195251, Russia}
\vspace*{0.225truein}
\pub{}
\vspace*{0.21truein}
\abstracts{
 We describe ${\rm U(N})$-monopoles ($N>1$) on Kerr black holes by the
parameters of the moduli space of
holomorphic vector ${\rm U(N)}$-bundles over ${\Bbb S}^2$ with the help of the
Grothendieck splitting theorem. For $N=2,3$ we obtain this
description in an explicit form as well as the estimates for the
corresponding monopole masses. This gives a possibility to adduce some
reasonings in favour of existence of both a {\it fine structure} for Kerr
black holes and the statistical ensemble tied with it which might generate
the Kerr black hole entropy.
}{}{}
\vspace*{1pt}\textlineskip 
\section{Introductory Remarks} 
\vspace*{-0.5pt}
\noindent

  The present paper is a natural continuation of our previous
work of Ref.\cite{Gon97}, so
we shall not dwell upon the motivation of studying the topics being
considered here so long as it has been done in Ref.\cite{Gon97}.
It should be here only noted
that one of the motivations of writing Ref.\cite{Gon97} was in the
Kerr black hole case to realize
the program performed in Refs.\cite{{Gon96},{Gon96p}} for the Schwarzschild
(SW) and Reissner-Nordstr{\"o}m (RN) black holes, namely, to try finding
the additional quantum numbers (nonclassical hair) characterizing Kerr black
holes that might help in building a statistical ensemble necessary to
generate the Kerr black hole entropy.

  The mentioned program for SW and RN black holes consisted in that with
the help of the classification of complex vector bundles over
${\Bbb S}^2$ and
the Grothendieck splitting theorem a number of
infinite series of ${\rm U(N)}$-magnetic monopoles at $N\geq1$ was constructed
in an explicit form on the SW and RN black holes.
Also the masses of the given monopoles were estimated to
show that they might reside in black holes as quantum objects. This gave
the possibility of applying to the problem of statistical substantiation of
the SW and RN black hole entropy.\cite{Gon96p}

The paper of Ref.\cite{Gon97} obtained some description of
${\rm U(1)}$-monopoles on
Kerr black holes. The present paper will be devoted to the extension
of the constructions of Ref.\cite{Gon97} to the ${\rm U(N)}$-monopoles ($N>1$)
on Kerr black holes along with an application to the problem of statistical
substantiation of the Kerr black hole entropy. In the present paper, however,
we shall use a gauge somewhat different from the gauge employed in
Ref.\cite{Gon97} to avoid unnecessary complications.

 In the Kerr black hole case we use the ordinary set of the local
Boyer-Lindquist coordinates
$t,r,\vartheta,\varphi$ covering the standard topology \bh of the
$4D$ black hole spacetimes except for a set of
the zero measure. At this the surface
$t=const., r=const.$ is an oblate ellipsoid with
topology ${\Bbb S}^2$ and the focal distance $a$ while $0\leq\vartheta<\pi,
0\leq\varphi<2\pi$. Under the circumstances we write
down the Kerr metric in the form
$$ds^2=g_{\mu\nu}dx^\mu\otimes dx^\nu\equiv(1-2Mr/\Sigma)dt^2-
\frac{\Sigma}{\Delta}dr^2-\Sigma d\vartheta^2-$$
$$[(r^2+a^2)^2-\Delta a^2\sin^2\vartheta]
\frac{\sin^2\vartheta}{\Sigma}d\varphi^2
+\frac{4Mra\sin^2\vartheta}{\Sigma}dtd\varphi     \eqno(1)$$
with $\Sigma=r^2+a^2\cos^2\vartheta$, $\Delta=r^2-2Mr+a^2$, $a=J/M$, where
$J, M$ are, respectively, a black hole mass and an angular moment.

  For inquiry we adduce the components of metric in the cotangent
bundle of manifold \bh with the metric (1) (in tangent bundle), so long as
we shall need them in calculations below. These are

$$ g^{tt}=\frac{1}{\Sigma\Delta}[(r^2+a^2)^2-\Delta a^2\sin^2\vartheta],
g^{rr}=-\frac{\Delta}{\Sigma}, g^{\vartheta\vartheta}=-\frac{1}{\Sigma},$$
$$g^{\varphi\varphi}=-\frac{1}{\Delta\sin^2\vartheta}(1-2Mr/\Sigma),
g^{t\varphi}=g^{\varphi t}=\frac{2Mra}{\Sigma\Delta}\>.  \eqno(2)$$

Besides we have $\delta=|\det(g_{\mu\nu})|=(\Sigma\sin\vartheta)^2$, $r_\pm=
M\pm\sqrt{M^2-a^2}$, so $r_+\leq r<\infty$, $0\leq\vartheta<\pi$,
$0\leq\varphi<2\pi$.

  Throughout the paper we employ the system of units with $\hbar=c=G=1$,
unless explicitly stated. Finally, we shall denote $L_2(F)$ the set of
the modulo square integrable complex functions on any manifold $F$
furnished with an integration measure.

\section{Description of U(N)-Monopoles}

In order to obtain the infinite families of
${\rm U(N)}$-monopoles for $N>1$, we should
use the Grothendieck splitting theorem\cite{{Okon80},{Grot56}} which asserts
that any complex vector bundle over ${\Bbb S}^2$ ( and, as a consequence, over
\bh) of rank $N>1$ [i. e., with the structural group
${\rm U(N)}$] is a direct sum of $N$ suitable complex line bundles over
${\Bbb S}^2$.
  The standard results of algebraic topology (see, e. g., Ref.\cite{Hus66})
say that
${\rm U(N)}$-bundles over ${\Bbb S}^2$ are in one-to-one correspondence with
elements of the fundamental group of ${\rm U(N)}$, $\pi_1[{\rm U(N)}]$. On the
other hand, in virtue
of the famous Bott periodicity\cite{Bot59} $\pi_1[{\rm U(N)}]={\Bbb{Z}}$
at $N\geq1$ and,
as a result, there exists the countable number of nontrivial complex vector
bundles of any rank $N>1$ over \bh. The sections of such bundles can be
qualified as topologically inequivalent configurations (TICs) of
$N$-dimensional (massless) complex scalar field. The above
classification confronts some $n\in{\Bbb{Z}}$ with each ${\rm U(N)}$-bundle
over \bh-topology. In what follows we shall call it the Chern number of the
corresponding bundle. TIC with $n=0$ can be called {\it untwisted} one
while the rest of the TICs with $n\not=0$ should be referrred to as
{\it twisted}.

  So far we tacitly implied that the ${\rm U(N)}$-bundles were supposed to be
differentiable. Really, they admit holomorphic structures and since each
differentiable complex line bundle over ${\Bbb S}^2$ admits only one
holomorphic structure (i. e., the holomorphic and differentiable
classifications of
complex line bundles over ${\Bbb S}^2$ coincide\cite{Okon80}) then
the Grothendieck
splitting theorem in fact gives a description of the moduli space
${\frak M}_N$ of
$N$-dimensional holomorphic complex vector bundles over
${\Bbb S}^2$. Namely, each
$N$-dimensional holomorphic complex vector bundle over
${\Bbb S}^2$ is defined by
the only $N$-plet of integers $(k_1,k_2,\ldots,k_N)\in{\Bbb{Z}}^N$,
$k_1\geq k_2\geq\ldots\geq k_N$. Two of such $N$-plets $(k_i)$ and
$(k_i^\prime)$ define the same
differentiable $N$-dimensional bundle if and only if $\sum\limits_i\,k_i=
\sum\limits_i\,k_i^\prime$.

  As was shown in Ref.\cite{Gon97}, each complex line bundle (with the
Chern number $k_i$, $i=1,2,...,N$) over \bh with
the metric (1) has a complete set of sections in
$L_2($\bh$)$, so using the fact that all the ${\rm U(N)}$-bundles over \bh can
be trivialized over the bundle chart of local coordinates ($t,r,\vartheta,
\varphi$) covering almost the whole manifold \bh, the mentioned set can be
written on the given chart in the form
$$f^{a\omega_i}_{k_il_im_i}=
\frac{1}{\sqrt{r^2+a^2}}e^{i\omega_it}R^{a\omega_i}_{k_il_im_i}(r)
Y_{k_il_im_i}(a\omega_i,\vartheta,\varphi)\,,$$
$$l_i=|k_i|,|k_i|+1,\ldots,|m_i|\leq l_i\,, \eqno(3)$$
where some properties of both the {\it monopole oblate
spheroidal harmonics}
$Y_{k_il_im_i}(a\omega_i,\vartheta,\varphi)$ and the eigenvalues
$\lambda_i=\lambda_{k_il_im_i}(a\omega_i)$
can be found in Ref.\cite{Gon97}, but we shall not need
them further.
As to the functions $R^{a\omega_i}_{k_il_im_i}(r)=R$
then, in the gauge under discussion, they obey the equation
$$\frac{d}{dr}\Delta\frac{d}{dr}\left(\frac{R}{\sqrt{r^2+a^2}}\right)+
\frac{(r^2+a^2)^2\omega_i^2-4Mm_ira\omega_i+m_i^2a^2}{\Delta}
\frac{R}{\sqrt{r^2+a^2}}=$$
$$-(\lambda_i+k^2_i)\frac{R}{\sqrt{r^2+a^2}}\>,   \eqno(4)$$
with $l_i=|k_i|, |k_i|+1,\ldots,|m_i|\leq l_i\,.$

  Now, in accordance with the Grothendieck splitting theorem, any section of
$N$-dimensional complex bundle $\xi_n$ over \bh with the Chern number
$n\in{\Bbb{Z}}$ can be represented by a $N$-plet ($\phi_1,\ldots,\phi_N$) of
complex scalar fields $\phi_i$, where each $\phi_i$ is a section of a complex
line bundle over \bh. According to the above, we can consider $\phi_i$ the
section of complex line bundle with the Chern number $k_i\in{\Bbb{Z}}$,
where the numbers $k_i$ are subject to the conditions
$$k_1\geq k_2\geq\ldots\geq k_N\>,$$
$$ k_1+k_2+\cdots+k_N=n\>.\eqno(5)$$
 As a consequence, we can require the $N$-plets
$(f^{a\omega_1}_{k_1l_1m_1},\ldots,f^{a\omega_N}_{k_Nl_Nm_N})$
to form the basis in $[L_2($\bh$)]^N$ for the sections of
$\xi_n$, $l_i=|k_i|,|k_i|+1,\ldots$, $|m_i|\leq l_i$,
and this will define the wave equation for a section $\phi=
(\phi_1,\ldots,\phi_N)$ of $\xi_n$ with respect to the metric (1)
$$\Biggl[I_N\Box-{1\over \Sigma^2\sin^2\vartheta}\times$$
{\footnotesize
$$\pmatrix{
2ik_1\cos\vartheta(a\sin^2\vartheta\partial_t+\partial_\varphi)
-k_1^2\cos^2\vartheta&0&\ldots&0\cr
0&2ik_2\cos\vartheta(a\sin^2\vartheta\partial_t+\partial_\varphi)
-k_2^2\cos^2\vartheta&\ldots&0\cr
\ldots&\ldots&\ldots&\ldots\cr
0&0&\ldots&2ik_N\cos\vartheta(a\sin^2\vartheta\partial_t+\partial_\varphi)
-k_N^2\cos^2\vartheta\cr}\Biggr]$$
}
$$\times\pmatrix{\phi_1\cr\phi_2\cr\vdots\cr\phi_N\cr}=0\>,
\eqno(6)$$
where $I_N$ is the unit matrix $N\times N$, $\Box =(\delta)^{-1/2}
\partial_\mu(g^{\mu\nu}(\delta)^{1/2}\partial_\nu)$ --- the conventional
wave operator conforming to metric (1).

  The Eq. (6) will, in turn, correspond to the lagrangian
$${\cal L}=\delta^{1/2}g^{\mu\nu}\overline{{\cal D_\mu}\phi}
{\cal D}_\nu\phi\>,\eqno(7)$$
 with
$\phi=(\phi_i)$ and a covariant derivative ${\cal D}_\mu=
\partial_\mu-igA^a_\mu\,T_a$ on sections of the bundle $\xi_n$, while the
overbar
in (7) signifies hermitian conjugation and the matrices $T_a$ will form
a basis of the Lie algebra of ${\rm U(N)}$ in $N$-dimensional space
(we, as is accepted in physics, consider the matrices $T^a$ hermitian),
$a=1,\ldots, N^2$, $g$ is a gauge coupling constant, i. e., we come
to a theory describing
the interaction of a $N$-dimensional twisted complex scalar field with the
gravitational field described by metric (1). The coefficients $A^a_\mu$ will
represent a connection in the given bundle $\xi_n$ and will describe some
nonabelian ${\rm U(N)}$-monopole.

  As can be seen, the Eq.(6) has the form ${\cal D^\mu}{\cal D}_
\mu\phi=0$, where ${\cal D}^\mu$ is a formal adjoint to
${\cal D}_\mu$ with regards to the scalar product induced by metric (1)
in $[L_2($\bh$)]^N$. That is, the operator ${\cal D}^\mu$ acts on the
differential forms $a_\mu dx^\mu$ with coefficients in the bundle $\xi_n$ in
accordance with the rule
$${\cal D}^\mu(a_\nu dx^\nu)=-{1\over\sqrt{\delta}}\partial_\mu(g^{\mu\nu}
\sqrt{\delta}a_\nu)
+ig\overline{A_\mu}g^{\mu\nu}a_\nu\>\eqno(8)$$
with $A_\mu=A^a_\mu T_a$.

 As a result, the equation ${\cal D}^\mu{\cal D}_\mu\phi=0$ takes the form
$$I_N\Box\phi-\frac{ig}{\sqrt{\delta}}\partial_\mu(g^{\mu\nu}\sqrt{\delta}
A_\nu\phi)-(ig\overline{A_\mu}g^{\mu\nu}\partial_\nu+
g^2g^{\mu\nu}\overline{A_\mu}A_\nu)\phi=0\>.\eqno(9)$$
 Comparing (6) with (9) gives a row of the (gauge) conditions:
$$A^a_rT_a=A^a_\vartheta T_a=0\,,\eqno(10)$$
$$g^{tt}A^a_tT_a+g^{t\varphi}A^a_\varphi T_a=
\frac{a\cos\vartheta}{g\Sigma}
\pmatrix{k_1&0&\ldots&0\cr
0&k_2&\ldots&0\cr
\ldots&\ldots&\ldots&\ldots\cr
0&0&\ldots&k_N\cr}\>,\eqno(11)$$
$$g^{\varphi t}A^a_tT_a+g^{\varphi\varphi}A^a_\varphi T_a=
\frac{\cos\vartheta}{g\Sigma\sin^2\vartheta}
\pmatrix{k_1&0&\ldots&0\cr
0&k_2&\ldots&0\cr
\ldots&\ldots&\ldots&\ldots\cr
0&0&\ldots&k_N\cr}\>,\eqno(12)$$
This gives
$$A^a_tT_a=\frac{a\cos\vartheta}{g\Sigma}
\pmatrix{k_1&0&\ldots&0\cr
0&k_2&\ldots&0\cr
\ldots&\ldots&\ldots&\ldots\cr
0&0&\ldots&k_N\cr}\>,\eqno(13)$$
$$A^a_\varphi T_a=-\frac{(r^2+a^2)\cos\vartheta}{g\Sigma}\>
\pmatrix{k_1&0&\ldots&0\cr
0&k_2&\ldots&0\cr
\ldots&\ldots&\ldots&\ldots\cr
0&0&\ldots&k_N\cr}\>
.\eqno(14)$$
Under the circumstances the connection in the bundle $\xi_n$ is
$A=A^a_\mu T_adx^\mu=
A^a_t(r,\vartheta)T_adt+
A^a_\varphi(r,\vartheta)T_ad\varphi$ which
yields the curvature matrix $F=dA+A\wedge A$ for $\xi_n$-bundle
in the form
$$F= F^a_{\mu\nu}T_a dx^\mu\wedge dx^\nu=
-\partial_r(A^a_tT_a)dt\wedge dr
-\partial_\vartheta(A^a_tT_a)dt\wedge d\vartheta$$
$$+\partial_r(A^a_\varphi T_a)dr\wedge d\varphi+
\partial_\vartheta(A^a_\varphi T_a)d\vartheta\wedge d\varphi
+[A^a_tT_a,A^b_\varphi T_b]dt\wedge d\varphi
\>, \eqno(15)$$
because the exterior differential $d=\partial_t dt+\partial_r dr+
\partial_\vartheta d\vartheta+\partial_\varphi d\varphi$ in coordinates
$t,r,\vartheta,\varphi$, while $[\cdot,\cdot]$ signifies the matrix commutator.
Then, with
taking into account Eqs. (13)--(14), we can see that the commutator in
the right-hand side of (15) vanish and
from here it follows that the first Chern class $c_1(\xi_n)$ of the bundle
$\xi_n$ can be chosen in the form
$$ c_1(\xi_n)=\frac{g}{4\pi}{\rm Tr}(F)
\>,\eqno(16)$$
so that, when integrating $c_1(\xi_n)$ over any surface 
$t=const., r=const.$,
we shall have with using (5) and (14)
$$ \int\limits_{S^2}\,c_1(\xi_n)=
\frac{g}{4\pi}\int\limits_{S^2}\,
{\rm Tr}[\partial_\vartheta(A^a_\varphi T_a)]d\vartheta\wedge d\varphi=-
\frac{n}{4\pi}\int\limits_{S^2}\,
\Omega\sin\vartheta d\vartheta\wedge d\varphi=$$
$$-\frac{n}{2}\int\limits_0^\pi\,\Omega\sin\vartheta d\vartheta=n
\>\eqno(17)$$
with
$$\Omega=\frac{(r^2+a^2)(a^2\cos^2\vartheta-r^2)}{\Sigma^2}\>,$$
which is equivalent to the conventional Dirac charge quantization condition
$$qg=4\pi n \> \eqno(18)$$
with (nonabelian) magnetic charge
$$q=\int\limits_{S^2}\,{\rm Tr}(F)\>.\eqno(19)$$
  Introducing the Hodge star operator $\ast$ conforming metric (1)
on 2-forms $F= F^a_{\mu\nu}T_adx^\mu\wedge dx^\nu$ with the values
in the Lie algebra of ${\rm U(N)}$ by the relation (see, e. g.,
Refs.\cite{Bes81})
$$(F^a_{\mu\nu} dx^\mu\wedge dx^\nu)\wedge
(\ast F^a_{\alpha\beta} dx^\alpha\wedge
dx^\beta)=(g^{\mu\alpha}g^{\nu\beta}-g^{\mu\beta}g^{\nu\alpha})
F^a_{\mu\nu}F^a_{\alpha\beta}
\sqrt{\delta}\,dx^0\wedge\cdots\wedge dx^3 \>,\eqno(20)$$
written in local coordinates $x^\mu$ [there is no summation over $a$
in (20)], in coordinates $t,r,\vartheta,\varphi$ we have for $F$ of (15)
$$\ast F=\ast F^a_{\mu\nu}T_a dx^\mu\wedge dx^\nu=$$
$$(g^{t\varphi}g^{\vartheta\vartheta}\frac{\partial A_t}{\partial\vartheta}
+g^{\vartheta\vartheta}g^{\varphi\varphi}
\frac{\partial A_\varphi}{\partial \vartheta})
\sqrt{|\delta|}\,dt\wedge dr
-(g^{\varphi t}g^{rr}\frac{\partial A_t}{\partial r}+
g^{rr}g^{\varphi\varphi}\frac{\partial A_\varphi}{\partial r})
\sqrt{|\delta|}\,dt\wedge d\vartheta$$
$$+(g^{tt}g^{\vartheta\vartheta}\frac{\partial A_t}{\partial\vartheta}+
g^{\vartheta\vartheta}g^{t\varphi}
\frac{\partial A_\varphi}{\partial \vartheta})\sqrt{|\delta|}\,dr\wedge d\varphi-
(g^{tt}g^{rr}\frac{\partial A_t}{\partial r}+g^{rr}g^{t\varphi}
\frac{\partial A_\varphi}{\partial r})\sqrt{|\delta|}\,d\vartheta\wedge d\varphi
                    \>\eqno(21)$$
with $A_t=A^a_tT_a$ and $A_\varphi=A^a_\varphi T_a$ of (13)--(14).
We can now consider the Yang-Mills equations
$$dF=F\wedge A - A\wedge F \>, \eqno(22)$$
$$d\ast F= \ast F\wedge A - A\wedge\ast F \>.\eqno(23)$$
It is clear that (22) is identically satisfied by the above $A, F$ --- this
is just the Bianchi identity holding true for any connection.\cite{Bes81}

 As for the Eq. (23), then, it is easy to check with the help of (13)--(14)
and (21) that $\ast F\wedge A=A\wedge\ast F$. Under this situation, from (21)
it follows that the
condition $d\ast F=0$ is equivalent to the equations
$$\frac{\partial}{\partial r}\left[\sqrt{|\delta|}\left(g^{rr}g^{\varphi t}
\frac{\partial A_t}{\partial r}+
g^{rr}g^{\varphi\varphi}\frac{\partial A_\varphi}{\partial r}\right)\right]
+\frac{\partial}{\partial \vartheta}\left[\sqrt{|\delta|}\left(
g^{t\varphi}g^{\vartheta\vartheta}\frac{\partial A_t}{\partial\vartheta}+
g^{\vartheta\vartheta}g^{\varphi\varphi}
\frac{\partial A_\varphi}{\partial \vartheta}\right)\right]=0\>,\eqno(24)$$

$$\frac{\partial}{\partial r}\left[\sqrt{|\delta|}\left(g^{tt}g^{rr}
\frac{\partial A_t}{\partial r}+g^{rr}g^{t\varphi}\frac{\partial A_\varphi}
{\partial r}\right)\right]
+\frac{\partial}{\partial \vartheta}\left[\sqrt{|\delta|}\left
(g^{tt}g^{\vartheta\vartheta}\frac{\partial A_t}{\partial\vartheta}+
g^{\vartheta\vartheta}g^{t\varphi}
\frac{\partial A_\varphi}{\partial \vartheta}\right)\right]=0\>.\eqno(25)$$
The direct evaluation with the aid of (13)--(14) shows that (24)--(25) are
satisfied. As a consequence, the Eq. (23) is fulfilled.

One can notice, moreover, that
$$Q_e=\int\limits_{S^2}\,{\rm Tr}(\ast F)=
-\int\limits_{S^2}\,g^{rr}{\rm Tr}\left(g^{tt}
\frac{\partial A_t}{\partial r}+g^{t\varphi}\frac{\partial A_\varphi}
{\partial r}\right)\sqrt{|\delta|}d\vartheta\wedge d\varphi=
-\frac{4\pi anr}{e}\int\limits_{-1}^1\,\frac{x\,dx}{\Sigma^2}=0
\>, \eqno(26)$$
where $x=\cos\vartheta$. As a result, an external observer does not see any
(internal) nonabelian electric charge $Q_e$ of the Kerr black hole for
any given $N$. Besides it should be emphasized that the total
(internal) nonabelian magnetic charge $Q_m$ of black hole which
should be considered as the one summed up over all the ${\rm U(N)}$-monopoles
for any given $N$ remains equal to zero because
$$Q_m=\frac{4\pi}{g}\sum\limits_{n\in{\Bbb{Z}}}\,n=0\>,\eqno(27)$$
so the external observer does not see any nonabelian magnetic charge of the
Kerr black hole either though ${\rm U(N)}$-monopoles are present on black
hole in the sense described above.

  To estimate the monopole masses we should use the $T_{00}$-component
of the energy-momentum tensor
$$T_{\mu\nu}=\frac{1}{4\pi}(-F^a_{\mu\alpha}F^a_{\nu\beta}g^{\alpha\beta}
+\frac{1}{4}
F^a_{\beta\gamma}F^a_{\alpha\delta}\,g^{\alpha\beta}g^{\gamma\delta}
g_{\mu\nu})\>.\eqno(28)$$
In our case
$$T_{00}=\frac{1}{4\pi}\{-g^{rr}(F^a_{tr})^2-
g^{\vartheta\vartheta}(F^a_{t\vartheta})^2+\frac{1}{4}g_{tt}[
g^{tt}g^{rr}(F^a_{tr})^2+g^{tt}g^{\vartheta\vartheta}(F^a_{t\vartheta})^2+$$
$$g^{rr}g^{\varphi\varphi}(F^a_{r\varphi})^2+
g^{\vartheta\vartheta}g^{\varphi\varphi}(F^a_{\vartheta\varphi})^2]\}
\>,    \eqno(29)$$
where $F^a_{tr}T_a=-\partial_r(A^a_tT_a)$,
$F^a_{t\vartheta}T_a=-\partial_\vartheta(A^a_tT_a)$,
$F^a_{r\varphi}T_a=\partial_r(A^a_\varphi T_a)$,
$F^a_{\vartheta\varphi}T_a=\partial_\vartheta(A^a_\varphi T_a)$.

Since we are in the asymptotically flat spacetime, we can calculate the
sought masses according to
$$m_{\rm mon}(k_1,\ldots,k_N)=\int\limits_{t=const}\,T_{00}\sqrt{\gamma}\,
dr\wedge d\vartheta\wedge d\varphi\>,   \eqno(30)$$
where
$$\sqrt{\gamma}=\sqrt{\det(\gamma_{ij})}=\sqrt{\Sigma/\Delta}\sin\vartheta
\sqrt{(r^2+a^2)^2-\Delta a^2\sin^2\vartheta}   \eqno(31)$$
for the metric $d\sigma^2=\gamma_{ij}dx^i\otimes dx^j$ on the hypersurface
$t=const$, while $T_{00}$ is computed at the given ${\rm U(N)}$-monopole.
Under the circumstances it is not complicated
to check that the leading term in asymptotic of
$T_{00}\sqrt{\gamma}$ at $r\to\infty$ will be defined by the addend
$g^{\vartheta\vartheta}g^{\varphi\varphi}(F^a_{\vartheta\varphi})^2$ of (29),
so one should solve the equation
$$F^a_{\vartheta\varphi}T_a=
\partial_\vartheta(A^a_\varphi T_a)\>,\eqno(32)$$
with $A^a_\varphi T_a$ of (14). Let us concretize it for $N=2,3$.

\section{Masses of U(2)- and U(3)-Monopoles}

  At $N=2$ we can take $T_1=I_2$, $T_a=\sigma_{a-1}$ at $a=2,3,4$, where
$\sigma_{a-1}$ are the ordinary Pauli matrices
$$\sigma_1=\pmatrix{0&1\cr 1&0\cr}\,,\sigma_2=\pmatrix{0&-i\cr i&0\cr}\,,
\sigma_3=\pmatrix{1&0\cr 0&-1\cr}\,\,. \eqno(33)$$
  Then the Eq. (32) gives $F^2_{\vartheta\varphi}=F^3_{\vartheta\varphi}=0$,
while
$$F^1_{\vartheta\varphi}=\frac{1}{2}(k_1+k_2)f(r,\vartheta)\>,
F^4_{\vartheta\varphi}=\frac{1}{2}(k_1-k_2)f(r,\vartheta)
  \eqno(34)$$
 with
$$f(r,\vartheta)=
-\partial_\vartheta\left[\frac{(r^2+a^2)\cos\vartheta}{g\Sigma}\right]
\>.\eqno(35)$$
This yields at $r\to\infty$
$$T_{00}\sqrt{\gamma}\sim
 \frac{\sin\vartheta}{64\pi g^2r^2}[(k_1+k_2)^2+(k_1-k_2)^2]\>.
   \eqno(36)$$
As a result, we can estimate (in usual units) according to (30)
$$m_{\rm mon}(k_1,k_2)\sim \left(\frac{\hbar^2c^2}{G}\right)
\frac{(k_1+k_2)^2+(k_1-k_2)^2}{16g^2}\int\limits_{r_+}^{\infty}
\,\frac{dr}{r^2}=
\frac{(k_1+k_2)^2+(k_1-k_2)^2}{16g^2r_+}
\left(\frac{\hbar^2c^2}{G}\right) \>.\eqno(37)$$

  At $N=3$ we can take $T_1=I_3$, $T_a=\lambda_{a-1}$
at $a=2,\ldots,9$, where $\lambda_{a-1}$ are the Gell-Mann matrices
$$\lambda_1=\pmatrix{0&1&0\cr 1&0&0\cr 0&0&0\cr}\,,
  \lambda_2=\pmatrix{0&-i&0\cr i&0&0\cr 0&0&0\cr}\,,
  \lambda_3=\pmatrix{1&0&0\cr 0&-1&0\cr 0&0&0\cr}\,,  $$
$$ \lambda_4=\pmatrix{0&0&1\cr 0&0&0\cr 1&0&0 \cr}\,,
   \lambda_5=\pmatrix{0&0&-i\cr 0&0&0\cr i&0&0\cr}\,,
   \lambda_6=\pmatrix{0&0&0\cr 0&0&1\cr 0&1&0\cr}\,, $$
$$\lambda_7=\pmatrix{0&0&0\cr 0&0&-i\cr 0&i&0\cr}\,,
  \lambda_8={1\over\sqrt3}\pmatrix{1&0&0\cr 0&1&0\cr
                   0&0&-2\cr}\,. \eqno(38) $$

  From (32) this yields $F^2_{\vartheta\varphi}=F^3_{\vartheta\varphi}=
F^5_{\vartheta\varphi}=F^6_{\vartheta\varphi}=
F^7_{\vartheta\varphi}=F^8_{\vartheta\varphi}=0$,
while
$$F^1_{\vartheta\varphi}=\frac{1}{3}(k_1+k_2+k_3)f(r,\vartheta)\>,
F^4_{\vartheta\varphi}=\frac{1}{2}(k_1-k_2)f(r,\vartheta)\>,
F^9_{\vartheta\varphi}=\frac{\sqrt{3}}{6}(k_1+k_2-2k_3)f(r,\vartheta)
  \eqno(39)$$
with $f(r,\vartheta)$ of (35). This gives
$$m_{\rm mon}(k_1,k_2,k_3)\sim
[(k_1+k_2+k_3)^2+\frac{9}{4}(k_1-k_2)^2+\frac{3}{4}(k_1+k_2-2k_3)^2]
\frac{1}{36g^2r_+}\left(\frac{\hbar^2c^2}{G}\right) \>.\eqno(40)$$
It is clear that the case of arbitrary $N$ can be treated analogously but
we shall not dwell upon it here. One can only noticed that the important case
is the one of ${\rm U(4)}$-monopoles because 4-dimensional complex vector 
bundles could describe TICs of both spinors and vector charged fields, i. e. 
these TICs physically could arise due to interaction with 
${\rm U(4)}$-monopoles. 
But this task requires its separate consideration.

Under the circumstances, evaluating the corresponding Compton wavelength
$\lambda_{\rm {mon}}(k_i)=\hbar/m_{\rm {mon}}(k_i)c$, we can see
that at any $n\ne0, N\geq1$, $\lambda_{\rm {mon}}(k_i)\ll r_g$, where
$r_g=r_+G/c^2$ is a gravitational radius of Kerr black hole, if
$g^2/\hbar c\ll1$. As a consequence, we come
to the conclusion that under certain conditions ${\rm U(N)}$-monopoles might
reside in black holes as quantum objects.

  So, we can see that the masses of ${\rm U(N)}$-monopoles really depend on
the parameters of the moduli space ${\frak M}_N$ of holomorphic vector bundles
over ${\Bbb S}^2$. Let us consider some possible issues for the 4$D$ Kerr 
black hole physics from this fact.

\section{Fine Structure of Kerr Black Hole
                               for Generating Its Entropy}

   Among the unsolved questions of modern $4D$ black hole physics the
so-called  black hole information problem admittedly ranks high. Referring
for more details, e. g., to Ref.\cite{Gon96p} (and references quoted
therein), it should be noted here that one aspect of the problem consists
in that
for an external observer any black hole looks like an object having in general
only a finite number of parameters (classical hair --- mass $M$, charge $Q$,
angular momentum $J$) and it is, therefore, unclear how these parameters
can encode
all the information about quantum particles of matter (which has been
collapsed to the black hole), particles that are being radiated \`a la Hawking.
As a consequence, it is impossible to distinguish all the black hole (pure)
states, so a black hole should, therefore, be described by a mixed state.
In other words, the system (black hole) has an entropy $S$ while
the latter does not correspond to any statistical ensemble, so long as
there is no infinite number of quantum (discrete) numbers connected with
this system to build an appropriate statistical ensemble.

 One can notice that recently the attempts have been undertaken to
statistically substantiate the entropy for a range of black holes derived
from string theory (see, e.g., Refs.\cite{SV96} and cited therein). These
black holes are, however, defined either in five dimensions or in four
dimensions they carry a row of not yet observable quantum numbers,
for example, the so-called axion charge. Therefore, such black holes cannot
be used to describe real astrophysical objects and can only serve as some model
examples. The real astrophysical objects having a claim on identifying with
black holes seem to be described by the (SW, RN and Kerr) solutions derived
from the standard Einstein gravity theory and we can call them
{\it classical} black holes. It is clear that this is the most physically
interesting set of black holes. But though for classical black holes also
one can point out a number of attempts on statistical substantiation of their
entropy, for example, within the framework of the so-called induced gravity
(see, e.g., Refs.\cite{FZ97} and quoted therein), after all, these efforts
have not yet led to any generally accepted statistical substantiation of the 
classical black hole
entropy either. As a result, searching for new approaches to this problem
for $4D$ classical black holes is well justified. In particular, in the above
attempts the global nontrivial topological properties of black holes were
practically ignored.

 But the results of Refs.\cite{{Gon96},{Gon96p}} for the SW and RN
black holes as well as the ones of both Ref.\cite{Gon97} and
the present paper for the Kerr black holes, however,
show that the natural candidates for additional quantum numbers
(nonclassical hair)
for classical black holes might be the quantum numbers parametrizing
${\rm U(N)}$-monopoles on black holes, so these numbers could be identified 
with ${\frak M}_N$. Really, as has been demonstrated recently in
Refs.\cite{{Gon97},{Gon95},{GF96}}
black holes can radiate \`a la Hawking for any TICs, for instance, of
complex scalar field with the Chern number $n\in {\Bbb Z}={\frak M}_1$ and
this occurs independently of other field configurations. More exact
analytical and numerical considerations\cite{GF96} show that, for instance,
in the SW black hole case, twisted TICs
can give the marked additional contribution of order 17 \% to the total
luminosity (summed up over all the TICs). This tells us that
there exists some {\it fine structure} in black hole physics which is
conditioned by nontrivial topological properties of black holes and the
given fine structure is able to markedly modify the black hole
characteristics, so long as, for example, the words " Hawking radiation for
complex scalar field " should be now understood as the radiation summed up
over all the TICs of complex scalar field on black hole. This, in turn,
leads to a marked increase of black hole luminosity.\cite{{Gon97},{GF96}}
In a sense, the black hole fine structure is quite analogous to the one of
atomic spectra in atomic physics where its existence enables us to achieve
an essentially better understanding of the whole structure of atoms.

Let us consider, therefore, more in detail in which way the above fine
structure might help to Kerr black holes
to form a statistical ensemble necessary to generate the Kerr black hole
entropy.

    As is known (see, e. g., Ref.\cite{Nov86}), the entropy $S$ of Kerr 
black hole can be introduced from
purely thermodynamical considerations and $S=\pi(r^2_++ a^2)$,
so when putting the internal energy of black
hole $U=M$, we obtain the temperature of black hole
$T={\partial U\over\partial S}=\frac{r_+-r_-}{8\pi Mr_+}$ through the standard
thermodynamical relation. It is obvious that $S$ corresponds to a formal
partition function
$$Z=\exp\left[-\frac{M}{T}+\pi(r^2_++a^2)\right]\>.\eqno(41)$$
The quantity $Z$ is formal because we cannot
point out any infinite statistical ensemble conforming to it, so that one
could obtain $Z$ by the usual Gibbs procedure, i. e., by averaging over
this ensemble.
   The results of Ref.\cite{Gon97} show that Kerr black hole can radiate
\`a la Hawking
for any TIC of complex scalar field with the Chern number
$n\in{\frak M}_1={\Bbb Z}$.
Such a radiation is practically
defined by a couple ($g_{\mu\nu}$, $n$) with the black hole metric
$g_{\mu\nu}$ of (1) and the
Chern number $n$ in the sense that these data are sufficient to describe the
physical quantities (for instance, luminosity $L(n)$) characterizing
the radiation process for TICs with the Chern number $n$.\cite{Gon97}
On the other hand,
as is known (see, e. g., Ref.\cite{Nov86}), the
Hawking effect is being obtained when considering the system (black hole +
matter field near it) semiclassically: the black hole is being described
classically while the matter field is being quantized. All mentioned above
suggests that the Hawking process occurs for the given pair
($g_{\mu\nu}$, $n$) when the black hole is in {\it a quantum state}
which can be
characterized by the {\it semiclassical} energy
$$E_n\sim M-\frac{\sqrt{M^2-a^2}}{4Mr_+}(r^2_++a^2)+{\cal E}(n)
            \eqno(42)$$
with ${\cal E}(n)\sim m_{\rm {mon}}(n)Tr_+\sim n^2T/4e^2$ with
$m_{\rm {mon}}(n)\sim n^2/4e^2r_+$ of Ref.\cite{Gon97},
$e=4.8\cdot10^{-10}\,{\rm cm^{3/2}\cdot g^{1/2}\cdot s^{-1}}$,
so long as ${\cal E}(n)$ is a natural
energy of the monopole with the Chern number $n$ residing in Kerr black hole,
since the additional contribution to the Hawking radiation is conditioned
actually by the same monopole.\cite{Gon97} We call $E_n$ semiclassical
because the first two terms of (42) in usual units does not depend on $\hbar$
while the third addend does (see Sec. 3).

  Under the circumstances there arises an infinite set of quantum states
($g_{\mu\nu}$, $n$) with the energy spectrum (42) for Kerr black hole.
After this, the Gibbs average takes the form
$$Z\sim\sum\limits_{n\in{\Bbb{Z}}}e^{-{E_n\over T}}=
\exp\left[-\frac{M}{T}+\pi(r^2_++a^2)\right]
\sum\limits_{n\in{\Bbb{Z}}}e^{-\frac{n^2}{4e^2}}=$$
$$\exp\left[-\frac{M}{T}+\pi(r^2_++a^2)\right]\vartheta_3(0,q)\eqno(43)$$
with the Jacobi theta function $\vartheta_3(v,q)$ and
$q=\exp\left(-\frac{1}{4e^2}\right)$. As a result, we obtain an inessential
constant additive correction $S_1=\ln\vartheta_3(0,q)$ independent of $M$ and
$a$ to the Kerr black hole entropy $S=\pi(r^2_++a^2)$ but now the latter is
the result of averaging over an infinite ensemble which should be considered
as inherent to Kerr black hole due to its nontrivial topological properties.

  It is clear that one can also consider all the triplets
($g_{\mu\nu}$, $k_1$, $k_2$),
where the pair ($k_1$, $k_2$) parametrizes the moduli space of
$U(2)$-monopoles ${\frak M}_2$, so that the Gibbs average should be
accomplished over
${\frak M}_2$ which will again lead to some inessential additional correction
to the entropy $S$ due to dependence (37). Moreover, this scheme will
obviosly hold true for ${\rm U(N)}$-monopoles at any $N>1$ if the Gibbs
average is accomplished over the moduli space ${\frak M}_N$.

\section{Concluding Remarks}
 The results of both the present paper and
Refs.\cite{{Gon97},{Gon96},{Gon96p},{Gon95},{GF96}} show that
the 4$D$ black hole physics can have a rich fine structure connected
with the topology \bh underlying the 4$D$ black hole spacetime manifolds.
It seems to be quite probable that this fine structure is tied with the
moduli spaces ${\frak M}_N$ of $N$-dimensional holomorphic vector bundles
over ${\Bbb S}^2$ and could manifest
itself in solving the whole number of problems within the $4D$ black hole
physics, so that one should seemingly thoroughly study the arising
possibilities, in particular, also in the Kerr-Newman metric case as a natural
charged generalization of Kerr metric.

On the other hand, the considerations of the present paper are actually of
the general interest for all the metrics (solutions of the Einstein equations)
which can naturally be realised on the topology \bh. To this class of metrics
one should, for example, attribute the Kottler metric, Taub---NUT metric, the
Vaidya metric (see, e. g., Ref.\cite{Exsol}). Especially, one should mark
the class of Tomimatsu-Sato metrics\cite{Tom72} and their charged
versions\cite{Ern73} which are natural extensions of Kerr and Kerr-Newman
metrics.

We hope to realise such a study elsewhere.

\nonumsection{Acknowledgements}
\noindent

    The work was supported in part by the Russian Foundation for
Basic Research (the grant no. 98-02-18380-a) and also
by GRASENAS (grant no. 6-18-1997).

\nonumsection{References}

\end{document}